\documentclass[prd,eqsecnum]{revtex4}

\newcommand{\lslash}[1]{#1\llap/}
\newcommand{\Eq}[1]{Eq.\ (\ref{#1})}
\newcommand{\Eqs}[2]{Eq.\ (\ref{#1}) and (\ref{#2})}
\newcommand{\gtilde}{\tilde g}

\newcommand{\Ref}[1]{Ref.\ \cite{#1}}

\begin{document}

\title{Electromagnetic properties of spin-3/2 Majorana particles}

\author{Jos\'e F. Nieves}
\email{nieves@ltp.uprrp.edu}
\affiliation{Laboratory of Theoretical Physics\\
Department of Physics, University of Puerto Rico \\
R\'{\i}o Piedras, Puerto Rico 00936}

\date{July 2013}

\begin{abstract}
The structure of the electromagnetic vertex function
of spin-3/2 particles is analyzed in a general way,
for the diagonal and off-diagonal couplings, of
charged as well as neutral particles including the case of
self-conjugate (Majorana) particles.
The restrictions imposed by common principles such as
electromagnetic gauge invariance and hermiticity are studied,
and the implications due to the discrete space-time symmetries or the
Majorana condition are deduced when they are applicable. In some cases
certain features of the vertex function are analogous to the
known ones for the spin-1/2 or spin-1 particles, but we find and discuss other
particular properties which are related to the spin-3/2
Rarita-Schwinger representation. For example, in the diagonal Majorana case,
the vertex function can contain a term of the form $\gamma_\mu\gamma_5$,
which resembles the axial charge radius term for Majorana neutrinos,
plus another one that resembles the vertex function for self-conjugate
spin-1 particles, but with the particularity that the two terms
may not appear independently of each other, but instead
with a specific relative coefficient. 
In essence this is due to the requirement that
the vertex function does not mix the genuine spin-3/2 degrees of freedom
with the spurious spin-1/2 components of the Rarita-Schwinger representation.
The analogous results for the other cases (off-diagonal and charged couplings)
are discussed as well.
\end{abstract}
\maketitle

\section{Introduction and Summary}

Since electrically neutral particles do not couple to the photon
at the tree-level, their electromagnetic properties can provide a window
to higher order effects or sectors of the Standard Model (SM)
and its extensions that may be difficult or impossible to probe directly.
For example, it has been known for a long time that Majorana neutrinos
can have neither electric nor magnetic dipole moments; it can only
have the so-called axial charge radius. Results of this type can also
be deduced, for example, for the transition moments between two
different Majorana neutrinos\cite{n:nuem,k:nuem,s:nuem,kg:nuem,k:nucpt},
and similar results hold for spin-1 particles\cite{np:spin1}
and for Majorana particles of arbitrary spin\cite{boudjemahanzaoui}.
By the same token, if departures from these results were observed
experimentally, it would have implications for some important principles
such as Lorentz and gauge invariance, $CPT$ and crossing symmetry.

There has been considerable activity recently in the study
of the effects of gravitinos in several cosmological
contexts such as nucleosynthesis and
inflation\cite{inflation1,inflation2,cosmology1,nucleosynthesis1},
as well as some areas of current phenomenological interest such as
dark matter\cite{darkmatter1,darkmatter2,darkmatter3}.
Many of these effects have to do with the electromagnetic properties
of gravitinos. While the electromagnetic properties of neutrinos
have been well studied, for example in the references mentioned,
and for the spin-1 bosons they were studied in a general
and comprehensive way in \cite{np:spin1}, for the spin-3/2 fermions
the existing studies are geared toward specific cases and applications,
such as the diagonal or transition vertices for charged baryons,
for example\cite{baryons,qcd}.
Thus it seems useful to look in a general way at the electromagnetic
properties of spin-3/2 particles, to provide a systematic
study that while being helpful for considering questions of fundamental
and intrinsic interest, is also useful for phenomenological applications
in the contexts mentioned. This work can also serve as a guide when
testing the SM and its extensions, the possible departures from them
or the breaking of some fundamental principles such as Lorentz or gauge
invariance, and can also be relevant in other problems involving
spin-3/2 particles such as spin-3/2 quark searches at the LHC\cite{quarks}
and others\cite{delgado1,delgado2}. In addition, although
we limit ourselves here to the spin-3/2 to spin-3/2 electromagnetic vertex,
it can also serve as a guide to consider the gravitino-neutrino
radiative transition, which has also been of interest in the context of dark
matter\cite{decay1,decay2,decay3,decay4,decay5,decay6,decay7,decay8,
decay9,decay10}.

Before embarking in the details we summarize the
strategy that we follow and the main results.
We denote by $j_\mu(Q,q)$ the matrix element of the electromagnetic
current operator $J^{\rm (em)}_\mu(0)$ between two spin-3/2 particles
states of momentum $k$ and $k^\prime$,
\begin{eqnarray}
\label{defmatrixelement}
j_\mu(Q,q) \equiv \langle f^\prime(k^\prime)|J^{\rm (em)}_\mu(0)|f(k)\rangle\,,
\end{eqnarray}
where it is convenient to introduce the variables
\begin{eqnarray}
q = k - k^\prime \,,\nonumber\\
Q = k + k^\prime\,,
\end{eqnarray}
and consider that matrix element as a function of them, as indicated
in \Eq{defmatrixelement}. The electromagnetic off-shell vertex function for
spin-3/2 particles, $\Gamma_{\alpha\beta\mu}(Q,q)$, is defined such that
\begin{eqnarray}
\label{jGammadef}
j_\mu(Q,q) = \bar U^{\prime\alpha}(k^\prime)
\Gamma_{\alpha\beta\mu}(Q,q)
U^\beta(k) \,,
\end{eqnarray}
where $U^\alpha(k)$ is a Rarita-Schwinger (RS) spinor.
The objective is to find the most general form of the vertex function
consistent with Lorentz and electromagnetic gauge invariance,
and other requirements that may be applicable such as the discrete
space-time symmetries or the Majorana nature of the particles.

It is useful to remember that when $f$ and $f^\prime$ represent electrically
neutral particles, then electromagnetic gauge invariance implies that
\begin{eqnarray}
\label{wardneutral}
q^\mu \Gamma_{\alpha\beta\mu}(Q,q) = 0\,,
\end{eqnarray}
for any values of $Q$ and $q$. On the other hand, if they are charged,
the relation analogous to \Eq{wardneutral} contains additional terms
on the right-hand side involving the inverse propagators of the
particles, reminiscent of the Ward identity in QED. Thus in that case,
instead of \Eq{wardneutral}, we get the weaker condition
for the on-shell matrix element
\begin{eqnarray}
\label{wardonshell}
q^\mu j_\mu(Q,q) = 0 \,.
\end{eqnarray}
In this paper we restrict ourselves to the on-shell matrix element
$j_\mu(Q,q)$, and therefore the condition as written in \Eq{wardonshell}
applies uniformly for all the cases.

We can summarize our main result as follows.
For the particular case in which the initial and final particle is the
same and it is self-conjugate (the \emph{diagonal Majorana case}),
we find that the vertex function
involves at most two independent terms, which can be written in the form
\begin{eqnarray}
\label{majdiagvertex1}
\Gamma_{\alpha\beta\mu} & = & m F
g_{\alpha\beta}\left(q^2\gamma_\mu - q_\mu\lslash{q}\right)\gamma_5
+ iF\left(q_\beta\epsilon_{\alpha\mu\nu\lambda} -
q_\alpha\epsilon_{\beta\mu\nu\lambda}\right)q^\nu Q^\lambda
\nonumber\\
&&\mbox{} + iG\left[
q^2 (g_{\mu\alpha} q_\beta + g_{\mu\beta} q_\alpha) -
2q_\mu q_\alpha q_\beta\right] \,,
\end{eqnarray}
where $m$ is the mass and the two corresponding form factors,
denoted here by $F$ and $G$, are real.
The term proportional to $\gamma_\mu\gamma_5$ is reminiscent
of the axial charge radius term for Majorana neutrinos\cite{n:nuem},
while the other two terms resemble the result that was obtained
in Ref.\ \cite{np:spin1} for the electromagnetic vertex function of
self-conjugate spin-1 particles (Eq. (4.14) in that reference).
An expression similar to \Eq{majdiagvertex1} was obtained in
\Ref{boudjemahanzaoui} for the vertex function in this same case.
However, the noteworthy feature of our result quoted above,
is the fact that the coefficients of the first two terms in \Eq{majdiagvertex1}
are not independent.  Further conditions exist if some discrete symmetries hold.
For example, if $CP$ holds, then $G = 0$, so that only the $F$ terms
in \Eq{majdiagvertex1} can be present. But if the $\gamma_\mu \gamma_5$
is present, then it must be accompanied by the second term in
\Eq{majdiagvertex1}. In essence this result is due to the requirement that
the vertex function does not mix the genuine spin-3/2 degrees of freedom
with the spurious spin-1/2 components of the RS representation.
We also obtain and discuss the corresponding results for the
off-diagonal and for the charged particle cases.

Our plan in the rest of the paper is then as follows.
In Section\ \ref{s:generalform} we write down the most general form
of the on-shell vertex function that is consistent with electromagnetic
gauge invariance in terms of a set of form factors that we denote as $a_i,b_i$.
The resulting representation of the vertex function
has the virtue that it involves simple
combinations of the $\gamma$ matrices, the momentum vectors,
the metric and the Levi-Civita tensors, which makes it amenable for
practical calculations of transition rates. However, as we point out there,
the coefficients $a_i,b_i$ so introduced must satisfy some relations,
related to the use of the RS spinor representation, which are difficult
to elucidate in general, and therefore that form of the vertex function
is not the most convenient form for studying
the implications of the various discrete symmetries and
making the connection with the diagonal or transition electromagnetic moments
of the particles. In Section\ \ref{s:physicalparametrization}
we write down another expression for the on-shell vertex function,
in terms of two matrices which we denote by $R_\mu, P_\mu$, and their products,
which by construction satisfy the physical requirements related to gauge
invariance and the use of the RS spinors that the vertex
function must satisfy, without having to impose further conditions. 
As we show there, the various terms in this second representation of the
vertex function have a simple interpretation in terms of the electromagnetic
moments of the spin-3/2 particles. The formulas that express the relations
between the two sets of form factors are given explicitly
there as well, and some details of the derivation are given in the appendix.
In Section\ \ref{s:discrete} we
study the implications due to the discrete transformations, such
as the $C, P, T$ transformations and their products, and
the conditions implied by the hermiticity of the interaction Lagrangian
and crossing symmetry. We consider the diagonal and off-diagonal (transition)
matrix element, and we include both the neutral and charged cases.
We then consider in detail the Majorana case, for which some of the
main results are summarized above, and in Section\ \ref{s:discussion}
we discuss the corresponding results for the off-diagonal and for the
charged particle cases.

\section{Parametrization of the vertex function}
\label{s:generalform}

\subsection{General form}
\label{subsec:generalabform}

The goal is to find the most general linearly independent set of
tensors matrices $\Gamma^{(A)}_{\alpha\beta\mu}(Q,q)$, such that we
can write
\begin{eqnarray}
\Gamma_{\alpha\beta\mu}(Q,q) = \sum_A
F_A \, \Gamma^{(A)}_{\alpha\beta\mu}(Q,q) \,,
\end{eqnarray}
in a way that \Eq{wardonshell} is satisfied.
The set of tensors $\Gamma^{(A)}_{\alpha\beta\mu}$
can be divided into two groups, according
to whether they contain an even or odd number of powers of $q$
(equivalently, whether they are even or odd under $q \rightarrow -q$).
Since the $F_A$ are functions only of $q^2$, 
this implies, corresponding to that classification, that
$\Gamma_{\alpha\beta\mu}(Q,q)$ can then be expressed in the form
\begin{equation}
\Gamma_{\alpha\beta\mu}(Q,q) = X_{\alpha\beta\mu} +
Y_{\alpha\beta\mu\nu}q^\nu\,,
\end{equation}
where both $X$ and $Y$ are even under $q \rightarrow -q$.
The transversality condition on $j_\mu$ implies that both pieces must
be, separately, transverse; i.e., 
\begin{eqnarray}
q^\mu
U^{\prime\alpha}(k^\prime) X_{\alpha\beta\mu} U^\beta(k) = 0 \,,\nonumber\\ 
q^\mu q^\nu 
U^{\prime\alpha}(k^\prime) Y_{\alpha\beta\mu\nu} U^\beta(k) = 0 \,.
\end{eqnarray}
The first condition implies that $X_{\alpha\beta\mu}$ must be of the form
\begin{eqnarray}
\label{XAdef}
X_{\alpha\beta\mu} = \tilde g_\mu{}^\nu A_{\alpha\beta\nu}\,,
\end{eqnarray}
where
\begin{eqnarray}
\label{defgtilde}
\gtilde_{\mu\nu} \equiv g_{\mu\nu} - \frac{q_\mu q_\nu}{q^2} \,.
\end{eqnarray}
The second condition is solved by taking
$Y_{\alpha\beta\mu\nu}$ to be the most general antisymmetric tensor
in the indices $\mu,\nu$. We can choose to express $Y_{\alpha\beta\mu\nu}$
in terms of its dual and write
\begin{eqnarray}
Y_{\alpha\beta\mu\nu} = 
{\epsilon_{\mu\nu}}^{\lambda\rho} B_{\alpha\beta\lambda\rho} \,.
\end{eqnarray}
In this way we obtain that we can express the vertex function
in the form
\begin{eqnarray}
\Gamma_{\alpha\beta\mu} =
\tilde {g_\mu}^\nu A_{\alpha\beta\nu} +
{\epsilon_{\mu\nu}}^{\lambda\rho} q^\nu B_{\alpha\beta\lambda\rho}\,,
\end{eqnarray}

The next step is to enumerate the possible tensor structures
that can appear in $A_{\alpha\beta\nu}$ and
$B_{\alpha\beta\lambda\rho}$. In doing so it is important to keep in
mind that the spinors satisfy the Dirac equation as well as the
auxiliary RS conditions
\begin{eqnarray}
\label{rsconditions}
k^\alpha U_\alpha(k) & = & 0 \,,\nonumber\\
\gamma^\alpha U_\alpha(k) & = & 0 \,,
\end{eqnarray}
with analogous relations for $U^\prime_\alpha(k^\prime)$.
The relations in \Eq{rsconditions} imply in particular the following,
\begin{itemize}
\item Terms containing a factor of $\gamma_\alpha$ or $\gamma_\beta$
do not appear.
\item We need not consider any term involving the factors
$\lslash{q}$ or $\lslash{Q}$ since any such factor can be
eliminated by means of the Dirac equation satisfied by the $U$ and
$U^\prime$ spinors (together with judicious use of the
anticommutation relations of the gamma matrices).
\item The terms that contain a factor of $q_\alpha$ ($q_\beta$)
are not independent of the analogous terms with $Q_\alpha$ ($Q_\beta$).
\item When writing $A_{\alpha\beta\nu}$, the terms involving $q_\nu$ disappear
when contracted with $\gtilde^{\mu\nu}$. 
\end{itemize}
In this way we arrive at the following terms for $A_{\alpha\beta\nu}$,
\begin{eqnarray*}
A_{\alpha\beta\nu} & = &
g_{\alpha\beta} \gamma_\nu (a_1  + a^\prime_1\gamma_5) +
g_{\alpha\beta} Q_\nu (a_2 + a^\prime_2\gamma_5) +
Q_\alpha Q_{\beta} \gamma_\nu (a_3 + a^\prime_3\gamma_5) +
Q_\alpha Q_{\beta} Q_\nu(a_4 + a^\prime_4\gamma_5)\nonumber\\
&&\mbox{} +
(g_{\nu\alpha} Q_{\beta} + g_{\nu\beta} Q_{\alpha})
(a_5 + a^\prime_5\gamma_5) +
(g_{\nu\alpha} Q_{\beta} - g_{\nu\beta} Q_{\alpha})
(a_6 + a^\prime_6\gamma_5)
\end{eqnarray*}
and proceeding in a similar way for $B_{\alpha\beta\lambda\rho}$,
\begin{eqnarray*}
B_{\alpha\beta\lambda\rho} & = &
b_1 g_{\alpha\lambda}g_{\beta\rho}+
b_2 g_{\alpha\lambda} Q_{\beta} Q_\rho +
b_3 g_{\beta\lambda} Q_{\alpha} Q_\rho \,.
\end{eqnarray*}
In the discussions in the next sections, it will be more convenient
to eliminate the appearances of $Q_{\alpha,\beta}$ in favor
of $q_{\alpha,\beta}$. Thus the final expressions for
$X_{\alpha\beta\mu}$ and $Y_{\alpha\beta\mu\nu} q^\nu$ are
\begin{eqnarray}
\label{ab}
X_{\alpha\beta\mu} & = &
g_{\alpha\beta} \tilde \gamma_\mu (a_1  + a^\prime_1\gamma_5) +
g_{\alpha\beta} \tilde Q_\mu (a_2 + a^\prime_2\gamma_5) +
q_\alpha q_{\beta} \tilde\gamma_\mu (a_3 + a^\prime_3\gamma_5) +
q_\alpha q_{\beta} \tilde Q_\mu(a_4 + a^\prime_4\gamma_5)\nonumber\\
&&\mbox{} +
(\tilde g_{\mu\alpha} q_{\beta} + \tilde g_{\mu\beta} q_{\alpha})
(a_5 + a^\prime_5\gamma_5) +
(\tilde g_{\mu\alpha} q_{\beta} - \tilde g_{\mu\beta} q_{\alpha})
 (a_6 + a^\prime_6\gamma_5)\,,\nonumber\\
Y_{\alpha\beta\mu\nu} q^\nu & = &
b_1 \epsilon_{\alpha\beta\mu\nu} q^\nu +
b_2 q_\beta\epsilon_{\alpha\mu\nu\lambda} q^\nu Q^\lambda +
b_3 q_\alpha\epsilon_{\beta\mu\nu\lambda} q^\nu Q^\lambda\,,
\end{eqnarray}
where
\begin{eqnarray}
\label{gammaQtilde}
\tilde\gamma_\mu & = & \tilde g_{\mu\nu} \gamma^\nu \,, \nonumber\\
\tilde Q_\mu & = & \tilde g_{\mu\nu} Q^\nu \,.
\end{eqnarray}
Notice that we have not included the terms of the form
\begin{eqnarray*}
\gamma_\mu \epsilon_{\alpha\beta\lambda\rho} q^\lambda Q^\rho\\
Q_\mu \epsilon_{\alpha\beta\lambda\rho} q^\lambda Q^\rho \\
\tilde {g_{\mu}}^\nu\epsilon_{\alpha\beta\nu\lambda} Q^\lambda\,.
\end{eqnarray*}
Using the identity
\begin{equation}
\label{5eidentity}
g_{\nu\mu}\epsilon_{\alpha\beta\lambda\rho} -
g_{\nu\alpha}\epsilon_{\mu\beta\lambda\rho} -
g_{\nu\beta}\epsilon_{\alpha\mu\lambda\rho} -
g_{\nu\lambda}\epsilon_{\alpha\beta\mu\rho} -
g_{\nu\rho}\epsilon_{\alpha\beta\lambda\mu} = 0\,,
\end{equation}
contracted with $Q^\nu q^\lambda Q^\rho$, $\gamma^\nu q^\lambda Q^\rho$
or $q^\nu q^\lambda Q^\rho$, those three terms
can be expressed in terms of the ones we have already included.
For similar reasons we have not included the terms involving gamma matrices
of the form $\varepsilon_{\gamma\delta\lambda\rho}\gamma^\lambda\gamma^\rho$
or $\varepsilon_{\gamma\delta\lambda\rho}\gamma^\rho$. The former
terms can be rewritten in terms of $\sigma_{\gamma\delta}\gamma_5$
(without the epsilon tensor)
and therefore can be absorbed in the $X$ terms that we have included above.
The terms in which the epsilon tensor is contracted
with one Dirac gamma matrix can be reduced by means of the
identity\footnote{We use the convention
$\gamma_5 = i\gamma^0 \gamma^1 \gamma^2 \gamma^3$ and
$\epsilon^{0123} = +1$} 
\begin{eqnarray}
i\epsilon_{\alpha\beta\gamma\lambda} \gamma^\lambda = 
\gamma_\alpha\gamma_\beta\gamma_\gamma\gamma_5 -
(g_{\alpha\beta}\gamma_\gamma - g_{\alpha\gamma}\gamma_\beta
+ g_{\beta\gamma}\gamma_\alpha)\gamma_5\,.
\end{eqnarray}
The term with the three gamma matrices either yields zero
when one gamma matrix is contracted with one of the spinors,
or it can be reduced, by means the Dirac equation when a gamma matrix
is contracted with $q$ or $Q$, to terms with
only one Dirac gamma matrix (times $\gamma_5$), which we already
included in the $X$ terms.

The form factors that appear in \Eq{ab} must
actually satisfy some conditions that follow from physical
requirements. However, those conditions depend on whether
we are considering charged or neutral particles, and also if
we are considering the diagonal or the off-diagonal (transition) vertex.
One set of conditions follows from the requirement
that there should not be any kinematic singularities at $q = 0$.
Another follows from the fact that we are dealing with a spin-3/2 particle.
Both imply that there are certain kinematic relations between the form
factors, none of which are taken into account in \Eq{ab}. We
now derive both sets of conditions.

\subsection{Kinematic singularities at $q = 0$}
\label{subsec:qzero}

The singular terms are
\begin{eqnarray}
\Gamma^{\rm (singular)}_{\alpha\beta\mu} & = & \frac{1}{q^2}\left[
g_{\alpha\beta} q_\mu \lslash{q} (a_1 + a^\prime_1\gamma_5) +
g_{\alpha\beta} q_\mu Q\cdot q (a_2 + a^\prime_2\gamma_5)\right.
\nonumber\\
&&\mbox{} + \left.
q_{\alpha}q_{\beta} q_\mu \lslash{q}(a_3 + a^\prime_3\gamma_5) +
q_\alpha q_\beta q_\mu Q\cdot q (a_4 + a^\prime_4\gamma_5)\right.
\nonumber\\
&&\mbox{} + \left.
2 a_5 q_{\alpha}q_{\beta} q_\mu +
2 a^\prime_5 q_{\alpha}q_{\beta} q_\mu \gamma_5\right] \,.
\end{eqnarray}
The conditions that the vertex function satisfies at $q = 0$ are different
depending on whether it is zero at the tree level or not.
As an example, let us consider the diagonal case and specifically
an electrically neutral particle. In this case,
there is no tree-level electromagnetic coupling and the vertex function is the
result of the higher order corrections. Therefore, in all the relevant
Feynman diagrams, the external photon line is attached to an internal
line and the momentum integrations ensure that all such contributions
are finite as $q \rightarrow 0$. Remembering that $Q\cdot q = 0$ and
that $\lslash{q} = 0$ between the spinors,
the requirement that the singularity at $q^2 = 0$ is absent implies
that the various coefficients $a_i(q^2)$ must satisfy
\begin{equation}
\label{qzeroconditions}
a^\prime_{1,3,5}(0) = a_5(0) = 0\,.
\end{equation}
While we could consider the nondiagonal case as well as the
charged-particle cases in a similar way,
we do not proceed any further in this directions since, as we will
see in Section\ \ref{subsec:physicalparametrization}
we will propose a more general and natural method to incorporate
these requirements.

\subsection{Absence of spin-1/2 component}
\label{subsec:pollution}

At this point it is useful to recall that the generators
of the Lorentz group for the RS spinor are
\begin{eqnarray}
\label{spin32generators}
(\Sigma_{\mu\nu})_{\alpha\beta} = 
(S_{\mu\nu})_{\alpha\beta} +
\frac{1}{2}g_{\alpha\beta}\sigma_{\mu\nu}\,,
\end{eqnarray}
where
\begin{eqnarray}
\label{spin1generators}
(S_{\mu\nu})_{\alpha\beta} \equiv
i(g_{\mu\alpha} g_{\nu\beta} - g_{\mu\beta} g_{\nu\alpha})
\label{Smunu}
\end{eqnarray}
are the spin-1 generators. It will be useful to also define the duals
\begin{eqnarray}
(\tilde \Sigma_{\mu\nu})_{\alpha\beta} \equiv \frac{1}{2}
\epsilon_{\mu\nu\lambda\rho}(\Sigma^{\lambda\rho})_{\alpha\beta}\,.
\end{eqnarray}

Notice that the terms $a_6$ and $a^\prime_6$ in \Eq{ab}
involve the spin-1 generators $S_{\mu\nu}$. The spin-1/2 generators
will surface by rewriting the $\gamma_\mu$ terms (e.g., the $a_1$ term),
using the Gordon decomposition
\begin{eqnarray}
\label{gordondecompV}
2m\bar U^\alpha(k^\prime) \gamma_\mu U^\beta(k) =
U^\alpha(k^\prime)\left[Q_\mu -
i\sigma_{\mu\nu}q^\nu\right]U^\beta(k)\,.
\end{eqnarray}
However, without further considerations, that would result
in the spin-1 generators $S_{\mu\nu}$
and the spin-1/2 generators $\frac{1}{2}\sigma_{\mu\nu}$ appearing
separately in the vertex function, instead of the
combination given in \Eq{spin32generators}. In a physically sensible
theory of the spin-3/2 particle, that cannot happen since it
would lead to unitarity inconsistencies
due to this mixing with the unphysical components of the RS
field.

Therefore we demand that in a physically sensible theory the
form factors that appear in \Eq{ab} must
satisfy further relations in such a way that only the spin-3/2
generators appear and not the spin-1 and spin-1/2 separately.

As an illustration, let us consider the identification of the
dipole moment terms in the diagonal case.
In addition to \Eq{gordondecompV}, we will use
\begin{eqnarray}
\label{gordondecompA}
Q_\mu\bar U^\alpha(k^\prime) \gamma_5 U^\beta(k) =
\bar U^\alpha(k^\prime)i\sigma_{\mu\nu}q^\nu \gamma_5U^\beta(k)\,.
\end{eqnarray}
together with the identity
\begin{eqnarray}
\label{sigmaidentity}
\sigma_{\mu\nu}\gamma_5 = i\tilde\sigma_{\mu\nu}\,,
\end{eqnarray}
where
\begin{eqnarray}
\label{tildesigma}
\tilde\sigma_{\mu\nu} = \frac{1}{2}\epsilon_{\mu\nu\lambda\rho}
\sigma^{\lambda\rho} \,.
\end{eqnarray}
Then, using \Eq{gordondecompV} to rewrite the $a_1$ term, and
\Eqs{gordondecompA}{sigmaidentity} to rewrite the $a^\prime_2$ term,
we obtain
\begin{eqnarray}
X_{\alpha\beta\mu} & = &
a_0 g_{\alpha\beta} Q_\mu -
\left(\frac{a_1}{2m}\right)g_{\alpha\beta}\,i\sigma_{\mu\nu}q^\nu +
a^\prime_2 g_{\alpha\beta}
\epsilon_{\mu\nu\lambda\rho}q^\nu\sigma^{\lambda\rho}\nonumber\\
&&\mbox{} +
a_6(g_{\mu\alpha} g_{\beta\nu} - g_{\mu\beta} g_{\alpha\nu})q^\nu +
\mbox{OT}\,,
\end{eqnarray}
where we have defined
\begin{eqnarray}
a_0 = a_2 + \frac{a_1}{2m} \,,
\end{eqnarray}
and $\mbox{OT}$ stands for the \emph{other terms}, which are not relevant
for the present discussion.

This form suggests what must happen. Namely, the parameters $a_1$ and
$a_6$ must be related at $q \rightarrow 0$ such that those terms
combine into one term  proportional to the tensor matrix
$(\Sigma_{\mu\nu})_{\alpha\beta}$ defined above.
Similarly the parameters $a^\prime_2$ and $b_1$
must be related such that their corresponding terms combine into
one proportional to $(\tilde \Sigma_{\mu\nu})_{\alpha\beta}$.
The two combinations obtained in this way will be identified with
the magnetic and electric dipole terms, respectively.
Thus, we assert that the following relations must hold
\begin{eqnarray}
\label{nonpollutionexamplerelations}
a_6 & = & \frac{a_1}{m} + O(q^2)  \nonumber\\
a^\prime_2 & = & \frac{i}{2}b_1 + O(q^2)\,,
\end{eqnarray}
so that
\begin{eqnarray}
a_6(g_{\mu\alpha} g_{\beta\nu} - g_{\mu\beta} g_{\alpha\nu})q^\nu
-\frac{a_1}{2m} g_{\alpha\beta}\,i\sigma_{\mu\nu}q^\nu =
\frac{a_1}{m}(-i\Sigma_{\mu\nu})_{\alpha\beta} q^\nu + O(q^2)\,,\nonumber\\
-\frac{a^\prime_2}{2}
g_{\alpha\beta}\epsilon_{\mu\nu\lambda\rho} q^\nu \sigma^{\lambda\rho} +
\frac{b_1}{2} \epsilon_{\mu\nu\lambda\rho} q^\nu
(\delta^\lambda_\alpha \delta^\rho_\beta -
\delta^\lambda_\beta \delta^\rho_\alpha) = 
\frac{b_1}{2} (-i\tilde \Sigma_{\mu\nu})_{\alpha\beta} q^\nu + O(q^2)\,,
\end{eqnarray}
and therefore
\begin{eqnarray}
\label{dipolemomentexample}
\Gamma_{\alpha\beta\mu} & = &
a_0 g_{\alpha\beta} Q_\mu +
\frac{a_1}{m}(-i\Sigma_{\mu\nu})_{\alpha\beta} q^\nu + 
\frac{b_1}{2} (-i\tilde \Sigma_{\mu\nu})_{\alpha\beta} q^\nu + \mbox{OT}\,.
\end{eqnarray}
The identification of the electric charge and dipole moments
is obvious in this form, but it is evident that it is not
practical to find all the appropriate relations
between the coefficients in this way.

\subsection{Nonrelativistic limit}
\label{subsec:nrlimit}

To motivate our strategy for taking into account
all the issues mentioned in Sections\ \ref{subsec:qzero} and
\ref{subsec:pollution} it is useful to consider the
nonrelativistic limit.
In the case of the spin-1/2 particles, if we consider the set of all
the possible bilinears
$\omega^{\prime\dagger} \Gamma \omega$,
where $\omega$ and $\omega^\prime$ are the two-component spinors
corresponding to the initial and final particle states,
it is not possible construct any higher rank three-dimensional
tensors other than the vector $\omega^{\prime\dagger} \sigma_i \omega$.
This follows from the relations that the spin matrices
$s^i \equiv \frac{\sigma^i}{2}$ satisfy,
\begin{eqnarray}
\label{asym}
[s^i,s^j] & = & i\epsilon^{ijk}s^k \,,\\
\label{sym}
\{s^i,s^j\} & = & \frac{1}{2}\delta^{ij} \,,
\end{eqnarray}
which is of course a reflection of the fact that, apart
from the identity matrix, the Pauli matrices span the space of
$2\times 2$ matrices. Therefore, the particle can have only
electric and magnetic dipole moments.

For a spin-1 particle, the spin matrices
\begin{eqnarray}
(S^i)^{jk} = -i \epsilon^{ijk} \,.
\end{eqnarray}
satisfy \Eq{asym}, but not \Eq{sym}. Therefore, the antisymmetric products
do not yield linearly independent matrices, but
the symmetric combinations do. Thus in this case, in addition to the dipole
moments, the particle can have quadrupole moments involving terms of the form
$\xi^\dagger\{S^i,S^j\}\xi$,
where $\xi$ and $\xi^\prime$ stand for the space part of the
spin-1 polarization vectors, written as column matrices.
The spin-1 particle can have no higher moments because, together
with the $S^i$, we have already exhausted the number of
independent $3 \times 3$ matrices. This is reflected in the
fact that the symmetrized product of the spin-1 matrices is given by
\begin{eqnarray}
\label{S3}
\{S^i,S^j\} S^k + (i \leftrightarrow k) + (j \leftrightarrow k) = 
2\left(S^i \delta^{jk} + S^j \delta^{ki} + S^k\delta^{ij}\right) \,,
\end{eqnarray}
which is the generalization of the familiar relation $(S^i)^3 = S^i$.
 
This suggests that the generalization to the spin-3/2
is to consider the symmetrized products of the spin-3/2 matrices
\begin{eqnarray}
\label{3dSigma}
(\Sigma^i)^{jk} = \frac{1}{2}\sigma^i \delta^{jk} + (S^i)^{jk} \,,
\end{eqnarray}
which are the three-dimensional versions of the generators defined
in \Eq{spin32generators}. Then, as we have already seen
in \Eq{dipolemomentexample}, the bilinears involving
$\Sigma^i$ are related to the dipole moments, while the quadrupole and
octupole moments are associated with the symmetrized combinations of the
products $\Sigma^i \Sigma^j$ and $\Sigma^i \Sigma^j \Sigma^k$,
respectively. As we have already mentioned, this requires that
the coefficients that appear in \Eq{dipolemomentexample} be related
in such a way that the spin-1/2 and spin-1 spin matrices do not
appear separately, but rather in the combinations given by the $\Sigma^i$
and their symmetrized products.
A bonus of these requirements, is that the \emph{spin-1/2 mixing problem}
goes away as we now show.

First it is useful to point out the following relation,
\begin{equation}
\label{sigSigrelation}
\sigma^j (\Sigma^i)^{jk} = \frac{1}{2}\sigma^i \sigma^k \,,
\end{equation}
which is readily verified by using the definition of the
$(\Sigma^i)^{jk}$ together with
$\sigma^j\sigma^i = \sigma^i\sigma^j + 2i\epsilon^{ijk}\sigma^k$.
This relation is easily generalized, by induction for example,
to the product of any number of $\Sigma$ matrices,
\begin{equation}
\label{nsigSigrelation}
\sigma^j (\Sigma^{i_1}\Sigma^{i_2}\ldots\Sigma^{i_n})^{jk} =
\frac{1}{2^n}\sigma^{i_1}\sigma^{i_2}\ldots\sigma^{i_n} \sigma^k \,.
\end{equation}
Now consider the spinor $\phi^i$ of a spin-3/2 particle
in its rest frame, constructed as a linear combination
of products of the form
\begin{equation}
\phi^i \sim \xi^i \omega\,,
\end{equation}
where $\xi^i$ and $\omega$ stand for the spin-1 and spin-1/2 spinors
defined above. Such a spinor satisfies
\begin{equation}
\label{RSnonrel}
\sigma^i \phi^i = 0\,,
\end{equation}
which is just the nonrelativistic limit of the RS condition expressing
the fact that $\phi^i$ has no spin-1/2 component of the products
$\xi^i \omega$. \Eq{nsigSigrelation} then implies 
\begin{equation}
\sigma^j (\Sigma^{i_1}\Sigma^{i_2}\ldots\Sigma^{i_n})^{jk}\phi^k = 0\,,
\end{equation}
which means that the spinor $(\Sigma^i)^{jk} \phi^k$,
or in general $(\Sigma^{i_1}\Sigma^{i_2}\ldots\Sigma^{i_n})^{jk} \phi^k$
also satisfies \Eq{RSnonrel} and therefore it is guaranteed to represent
a spin-3/2 spinor without any spin-1/2 component.

This then yields the following recipe.
If we write the transition amplitude in the form
\begin{equation}
M = \phi^{\prime j\dagger} T^{jk} \phi^k \,.
\end{equation}
the above results imply that, if $T$ is constructed from the $(\Sigma^i)^{jk}$
and its products, then $T$ does not have matrix elements between
the spin-1/2 and spin-3/2 components. In particular, if
the initial spinor $\phi^i$ satisfies \Eq{RSnonrel},
then $T$ has matrix elements only if the final spinor also satisfies
that same condition; there is no spin-1/2 mixing problem.

The covariant version of this argument is based  on the
following identities which generalize the relations
given in \Eqs{sigSigrelation}{nsigSigrelation},
\begin{eqnarray}
\label{nsigSigrelationCov}
\gamma^\alpha \left(\Sigma_{\mu\nu}\right)_{\alpha\beta} = \frac{1}{2}
\sigma_{\mu\nu} \gamma_\beta \,,\nonumber\\
\gamma^\alpha \left(\Sigma_{\mu_1\nu_1} \cdots
\Sigma_{\mu_n\nu_n}\right)_{\alpha\beta} & = &
\frac1{2^n} \sigma_{\mu_1\nu_1} \cdots \sigma_{\mu_n\nu_n} \gamma_\beta \,,
\end{eqnarray}
with the understanding that the products are defined by
\begin{equation}
\left(\Sigma_{\mu_1\nu_1}\Sigma_{\mu_2\nu_2}\right)_{\alpha\beta} =
\left(\Sigma_{\mu_1\nu_1}\right)_{\alpha\gamma}
{\left(\Sigma_{\mu_2\nu_2}\right)^\gamma}_\beta \,.
\end{equation}
Denoting the transition amplitude by
$\bar U^{\prime\alpha}T_{\alpha\beta} U^\beta$,
these imply that if the initial spinor
satisfies the RS condition, then the final spinor $T_{\alpha\beta}
U^\beta$ also satisfies the RS condition if $T_{\alpha\beta}$ is
constructed from the $\Sigma_{\mu\nu}$ and their products.

In summary, the form factors that appear in the vertex function,
for example in \Eq{dipolemomentexample}, must be related such that
the spin-1/2 and spin-1 matrices do not appear separately
in the amplitude but only in the combinations given by $\Sigma_{\mu\nu}$
and their products. Specifically, apart from the electric monople term,
the vertex function must be expressible in terms of the following matrices
\begin{eqnarray}
\Sigma_{\mu\nu}\,, && \tilde\Sigma_{\mu\nu}\,, \nonumber\\
\Sigma_{\mu\nu}\Sigma_{\lambda\rho}\,, &&
\Sigma_{\mu\nu}\tilde\Sigma_{\lambda\rho}\,,\nonumber\\
\Sigma_{\mu\nu}\Sigma_{\lambda\rho}\Sigma_{\sigma\tau}\,, &&
\tilde\Sigma_{\mu\nu}\Sigma_{\lambda\rho}\Sigma_{\sigma\tau}\,,
\end{eqnarray}
This program is carried out in the next section.

\section{Physical parametrization of the vertex function} 
\label{s:physicalparametrization}

\subsection{Physical form factors}
\label{subsec:physicalparametrization}

It is convenient to define
\begin{eqnarray}
R_\mu & = & \Sigma_{\mu\nu} q^\nu \nonumber\\
P_\mu & = & \tilde\Sigma_{\mu\nu} q^\nu \,.
\end{eqnarray}
Then, remembering what we have discussed in the previous section,
an economical way to incorporate the conditions
that the vertex function must satisfy is to write it in the form
\begin{eqnarray}
\label{physicalparametrization}
\Gamma_{\alpha\beta\mu} & = & A_1 \tilde Q_\mu g_{\alpha\beta} +
i A_2 \tilde Q_\mu g_{\alpha\beta}\gamma_5
+ i B_1 (R_\mu)_{\alpha\beta} + i B_2 (P_\mu)_{\alpha\beta} \nonumber\\
&& \mbox{}
+ C_1 \tilde Q_\mu (R\cdot R)_{\alpha\beta} +
C_2 \tilde Q_\mu (R\cdot P)_{\alpha\beta} \nonumber\\
&& \mbox{} +
\Big(i C_3 [R_\mu, P_\nu ]_{\alpha\beta} +
C_4 \{ R_\mu, P_\nu \}_{\alpha\beta} +
C_5 \{ P_\mu, R_\nu \}_{\alpha\beta} \Big) Q^\nu \nonumber\\
&&\mbox{} +
i D_1 \{R_\mu, R\cdot R\}_{\alpha\beta} +
D_2 [R_\mu, R\cdot R]_{\alpha\beta} \nonumber\\
&& \mbox{} +
i E_1 \{R_\mu, R\cdot P\}_{\alpha\beta} +
i E_2 (R^\lambda P_\mu R_\lambda)_{\alpha\beta} \,,
\end{eqnarray}
where $\tilde Q_\mu$ is defined in \Eq{gammaQtilde}.
This form ensures that it satisfies the
requirements considered in Section\ \ref{subsec:pollution}.
The antisymmetric and symmetric combinations in terms of the
commutators and anticommutators have been chosen such that
the various terms have simple transformation properties under
the various discrete symmetries, as will be discussed below.

The tree-level electromagnetic coupling of the RS field contributes
only to $A_1$ and $B_1$. All the other terms in \Eq{physicalparametrization}
can arise only through the higher order corrections. Therefore, using
the argument stated in Section\ \ref{subsec:qzero},
the coefficients must satisfy
\begin{eqnarray}
A_2(0) = C_1(0) = C_2(0) = 0\,,
\end{eqnarray}
to reflect the absence of the kinematic singularity at $q^2 = 0$
in all such contributions.
In the nondiagonal case, or in the diagonal case in which the
particle is electrically neutral, a similar condition applies to
$A_1$ as well,
\begin{eqnarray}
A_1(0) = 0 \,,
\end{eqnarray}
but in the diagonal, electrically charged case, $A_1(0)$ is the electric
monopole term and it is of course nonzero.

Other combinations of $R, P, Q$ that could be written down are not
independent of the ones included in \Eq{physicalparametrization}.
For example, using the commutation relations of the generators of the
Lorentz group, it follows that
\begin{equation}
\label{RRcommutator}
i[R_\mu,R_\nu] = q_\mu R_\nu - q_\nu R_\mu + q^2 \Sigma_{\mu\nu} \,,
\end{equation}
while the same commutation relations also imply
\begin{equation}
i[\Sigma_{\mu\nu},\tilde \Sigma_{\alpha\beta}] = 
\epsilon_{\alpha\beta\mu\lambda} {\Sigma_\nu}^\lambda -
\epsilon_{\alpha\beta\nu\lambda} {\Sigma_\mu}^\lambda\,,
\end{equation}
which in turn leads to
\begin{equation}
\label{RPcommutator}
i[R_\mu,P_\nu] = \epsilon_{\mu\nu\lambda\rho} R^\lambda q^\rho \,. 
\end{equation}
In particular, the latter relation implies
\begin{eqnarray}
\label{RPcommutator2}
\left[R_\mu,P^\mu\right] & = & 0 \,,\nonumber\\
\left[P_\mu,R_\nu\right] & = & \left[R_\mu,P_\nu\right]\,.
\end{eqnarray}
which justifies why we do not include the terms involving
$P\cdot R$ and $\left[P_\mu,R_\nu\right]Q^\nu$ 
in \Eq{physicalparametrization}. The cubic terms
\begin{equation}
(P_\mu R\cdot R)_{\alpha\beta}\,,
(R\cdot R P_\mu )_{\alpha\beta}\,,
(P^\lambda R_\mu R_\lambda)_{\alpha\beta}\,,
(R^\lambda R_\mu P_\lambda)_{\alpha\beta}\,,
(R^\lambda R_\mu R_\lambda)_{\alpha\beta}\,,
\end{equation}
are redundant due to the relations
\begin{eqnarray}
\label{R2Prelations}
P_\mu R\cdot R = R\cdot R P_\mu & = & R^\lambda P_\mu R_\lambda + q^2 P_\mu \,,
\nonumber\\
R^\lambda R_\mu P_\lambda & = & R\cdot P R_\mu - q^2 P_\mu\,,
\nonumber\\
P^\lambda R_\mu R_\lambda & = & R_\mu R\cdot P - q^2 P_\mu\,,
\end{eqnarray}
which are easily derived from \Eq{RPcommutator} (and using \Eq{RRcommutator}),
and the relation
\begin{equation}
\label{RRmuR}
(R^\lambda R_\mu R_\lambda)_{\alpha\beta} =
\frac{1}{2}(R_\mu R\cdot R + R\cdot R R_\mu)_{\alpha\beta}
- q^2 (R_\mu)_{\alpha\beta}\,,
\end{equation}
which also follows from \Eq{RRcommutator}.

In some other cases, a given term becomes redundant when the RS
conditions are taken into account. For example, using the well-known
identities for the product of two Levi-Civita tensors, it is
straightforward to show that
\begin{equation} 
(P\cdot P)_{\alpha\beta} = (R\cdot R)_{\alpha\beta} -
\frac{q^2}{2}(\Sigma_{\mu\nu} \Sigma^{\mu\nu})_{\alpha\beta}\,.
\end{equation}
while, on the other hand, a little algebra shows that
\begin{equation}
(\Sigma_{\mu\nu} \Sigma^{\mu\nu})_{\alpha\beta} = 
9 g_{\alpha\beta} + 2i \sigma_{\alpha\beta}
\end{equation}
Thus, using the relation
\begin{equation}
\label{sigmag}
i\sigma_{\alpha\beta} = g_{\alpha\beta} - \gamma_\alpha \gamma_\beta\,,
\end{equation}
the RS auxiliary condition implies that the term
involving $(P\cdot P)_{\alpha\beta}$ does not yield a new contribution.
Similarly, as shown in \Eq{cubicterms}, the terms
$(R\cdot P R_\mu )_{\alpha\beta}$ and $(R_\mu R\cdot P)_{\alpha\beta}$
differ by a quantity that vanishes between the spinors and therefore
only their anticommutator ($E_1$) is included in \Eq{physicalparametrization}.
Incidentally, notice also that terms with explicit factors of $\gamma_5$
in \Eq{physicalparametrization} would be redundant.
For example, $R_\mu\gamma_5$ is contained in $R_\mu R\cdot P$, as
can be seen either from \Eq{cubicterms} or \Eq{cubics} below.

\Eq{physicalparametrization} can be
rewritten in terms of the form factors introduced in \Eq{ab} by
expanding the products of the $\Sigma$ matrices and following the rules
stated in Section\ \ref{subsec:generalabform}. As an example, let us consider
in some detail the $C_1$ term in \Eq{physicalparametrization}.
After a little bit of algebra, we find
\begin{eqnarray}
(\Sigma_{\mu\sigma})_{\alpha\lambda} {({\Sigma^\mu}_\tau)^\lambda}_\beta & = &
\frac{5}{4}g_{\sigma\tau} g_{\alpha\beta}
+ 3g_{\sigma\alpha} g_{\tau\beta} + g_{\tau\alpha} g_{\sigma\beta}
+ \frac{1}{2} \gamma_\sigma \gamma_\tau g_{\alpha\beta}
\nonumber\\
&&\mbox{} 
- \frac{1}{2} \gamma_\alpha \gamma_\sigma g_{\tau\beta}
- \frac{1}{2} \gamma_\sigma \gamma_\beta g_{\tau\alpha}
- \frac{1}{2} \gamma_\alpha \gamma_\tau g_{\sigma\beta}
- \frac{1}{2} \gamma_\tau \gamma_\beta g_{\sigma\alpha} \,,
\end{eqnarray}
and therefore
\begin{equation}
\label{RR}
(R\cdot R)_{\alpha\beta} = \frac{7}{4}q^2 g_{\alpha\beta}
+ 4q_\alpha q_\beta + L_{1\alpha\beta} \,,
\end{equation}
where
\begin{equation}
L_{1\alpha\beta} = 
-(q_\alpha \lslash{q}\gamma_\beta + q_\beta\gamma_\alpha\lslash{q}) \,.
\end{equation}
In a similar fashion we find for the term involving $C_2$
\begin{equation}
\label{RP}
(R\cdot P)_{\alpha\beta} = -\frac{5}{4}i q^2 g_{\alpha\beta} \gamma_5
+ L_{2\alpha\beta}
\end{equation}
where
\begin{equation}
L_{2\alpha\beta} = \frac{i}{2} q^2 \gamma_\alpha \gamma_\beta\gamma_5 \,.
\end{equation}

In all the cases, the terms that contain a factor of $L_{1\alpha\beta}$
or $L_{2\alpha\beta}$ drop out when they appear between the RS spinors
due to the RS condition and the use of the relation given
in \Eq{nsigSigrelationCov}. For example, consider a term of the form
\begin{equation}
\bar U^{\prime\alpha} (L_{1} O)_{\alpha\beta} U^\beta \,,
\end{equation}
where $O$ is any matrix built from products of $R_\mu$ and/or $P_\mu$.
The $\gamma_\alpha$ term of $L_{1}$ gives zero by the RS condition
satisfied by the $U^{\prime\alpha}$ spinor while the term with
$\gamma_\beta$ can be pushed all the way to the right-hand side by the
use of \Eq{nsigSigrelationCov}, which then yields zero by the RS
condition on the $U^{\beta}$ spinor. A similar argument holds for the
terms of the form $O L_{1}$ and the analogous terms with $L_1$
replaced by $L_2$, and we summarize this by writing
\begin{equation}
\label{OL}
\bar U^{\prime\alpha} (L_{1,2} O)_{\alpha\beta} U^\beta = 
\bar U^{\prime\alpha} (OL_{1,2})_{\alpha\beta} U^\beta = 0 \,.
\end{equation}
Therefore, any term that contains a factor of $L_1$ or $L_2$
will not contribute in the expression for the matrix element of
the vertex function. 

In addition the terms involving $\sigma_{\lambda\rho}$ must be reduced
using the Gordon identities,
\begin{eqnarray}
\label{Gordon}
\bar U^{\prime\alpha}
i\sigma_{\mu\nu} q^\nu
U^\beta & = & \bar U^{\prime\alpha}
(Q_\mu - M\gamma_\mu)
U^\beta =
\bar U^{\prime\alpha}
(\tilde Q_\mu - M \tilde \gamma_\mu)
U^\beta\,,\nonumber\\
\bar U^{\prime\alpha}
i\sigma_{\mu\nu} q^\nu \gamma_5
U^\beta & = &
\bar U^{\prime\alpha}
(Q_\mu + \Delta\gamma_\mu)\gamma_5\,
U^\beta =
\bar U^{\prime\alpha}
(\tilde Q_\mu + \Delta\tilde\gamma_\mu)\gamma_5\,
U^\beta \,,\nonumber\\
\bar U^{\prime\alpha}
i\sigma_{\mu\nu} Q^\nu
U^\beta & = &
\bar U^{\prime\alpha}
(q_\mu - \Delta \gamma_\mu)
U^\beta \,,\nonumber\\
\bar U^{\prime\alpha}
i\sigma_{\mu\nu} Q^\nu\gamma_5
U^\beta & = &
\bar U^{\prime\alpha}
(q_\mu + M\gamma_\mu)\gamma_5
U^\beta \,,
\end{eqnarray}
where $\tilde\gamma_\mu$ and $\tilde Q_\mu$ are defined in \Eq{gammaQtilde},
and
\begin{eqnarray}
M  =  m + m' \,, \qquad \Delta = m - m' \,,
\end{eqnarray}
with $m$ and $m^\prime$ denoting the mass of the initial and final
particle, respectively. Thus we can write
\begin{eqnarray}
\label{R}
\bar U^{\prime\alpha}
\left(R_\mu\right)_{\alpha\beta}
U^\beta & = & \bar U^{\prime\alpha}\left[
i\left(\tilde g_{\mu\alpha}
q_\beta - \tilde g_{\mu\beta} q_\alpha\right) -
\frac{i}{2}g_{\alpha\beta}\left(\tilde Q_\mu - M \tilde\gamma_\mu\right)\right]
U^\beta\,,\nonumber\\
\bar U^{\prime\alpha}
\left(P_\mu\right)_{\alpha\beta}
U^\beta & = & \bar U^{\prime\alpha}\left[
i\epsilon_{\alpha\beta\mu\nu}
q^\nu - \frac{1}{2}
g_{\alpha\beta}\left(\tilde Q_\mu + \Delta \tilde\gamma_\mu\right)\gamma_5
\right]U^\beta \,,
\end{eqnarray}
where we have used \Eq{equivalentPform} and also
\begin{eqnarray}
g_{\mu\alpha} q_\beta - g_{\mu\beta} q_\alpha
= \tilde g_{\mu\alpha} q_\beta - \tilde g_{\mu\beta} q_\alpha \,.
\end{eqnarray}
Using \Eqs{RR}{RP}, and remembering \Eq{OL}, we then obtain
\begin{eqnarray}
\bar U^{\prime\alpha}
(R\cdot R)_{\alpha\beta}
U^\beta & = &
\bar U^{\prime\alpha}\left[
\frac{7}{4}q^2 g_{\alpha\beta} + 4q_\alpha q_\beta
\right]U^\beta\,,\nonumber\\
\bar U^{\prime\alpha}
(R\cdot P)_{\alpha\beta} 
U^\beta & = &
\bar U^{\prime\alpha}\left[
-\frac{5}{4}i q^2 g_{\alpha\beta}\gamma_5
\right]U^\beta \,, 
\end{eqnarray}

The cubic terms in $R$ and $P$ as well as the terms involving the
commutator or anticommutator of $R$ and $P$ can be reduced in
similar fashion. The details are given in the appendix and the
final results are summarized below. First for the cubic terms
\begin{eqnarray}
\label{cubics}
\bar U^{\prime\alpha}
\{R_\mu, R\cdot R\}_{\alpha\beta}
U^\beta & = &
\bar U^{\prime\alpha}\left[
\frac{7}{2} q^2 (R_\mu)_{\alpha\beta} -
4i q_\alpha q_\beta (\tilde Q_\mu - M \tilde\gamma_\mu) +
4i q^2 (\tilde g_{\mu\alpha} q_\beta - \tilde g_{\mu\beta} q_\alpha)
\right]U^\beta\,,\nonumber\\
\bar U^{\prime\alpha}
[R_\mu, R\cdot R]_{\alpha\beta} 
U^\beta & = &
\bar U^{\prime\alpha}\left[
4i q^2 (\tilde g_{\mu\alpha} q_\beta + \tilde g_{\mu\beta} q_\alpha)
\right]U^\beta\,,\nonumber\\
\bar U^{\prime\alpha}
(R_\mu R \cdot P)_{\alpha\beta}
U^\beta & = &
\bar U^{\prime\alpha}\left[
-\frac{5}{8} q^2 g_{\alpha\beta}
(\tilde Q_\mu + \Delta\tilde\gamma_\mu)\gamma_5
+ \frac{5}{4} q^2(\tilde g_{\mu\alpha} q_\beta
- \tilde g_{\mu\beta} q_\alpha)\gamma_5
\right]U^\beta\,,\nonumber\\
\bar U^{\prime\alpha}
(R^\lambda P_\mu R_\lambda)_{\alpha\beta} 
U^\beta & = &
\bar U^{\prime\alpha}\left[
\frac{3}{4} q^2 (P_\mu)_{\alpha\beta} -
2 q_\alpha q_\beta (\tilde Q_\mu + \Delta \tilde\gamma_\mu)\gamma_5 
\right]U^\beta\,.
\end{eqnarray}
and finally for the terms involving the commutator or anticommutator
of $R$ and $P$,
\begin{eqnarray}
\label{RPcommanticommreduced}
\bar U^{\prime\alpha}
\left[R_\mu, P_\nu\right]_{\alpha\beta} Q^\nu
U^\beta & = &
\bar U^{\prime\alpha}\left[
q_\alpha\epsilon_{\beta\mu\nu\lambda}q^\nu Q^\lambda
- q_\beta\epsilon_{\alpha\mu\nu\lambda} q^\nu Q^\lambda
+ \frac{i}{2} g_{\alpha\beta}
\left(M(q^2 - \Delta^2) \tilde\gamma_\mu - \Delta M\tilde Q_\mu\right)\gamma_5
\right]U^\beta\,,\nonumber\\
\bar U^{\prime\alpha}
\left\{P_\mu, R_\nu\right\}_{\alpha\beta} Q^\nu
U^\beta & = &
\bar U^{\prime\alpha}\left[
-\frac{i}{2}q^2 g_{\alpha\beta} \tilde Q_\mu \gamma_5
- q_\alpha\epsilon_{\beta\mu\nu\lambda}q^\nu Q^\lambda
- q_\beta\epsilon_{\alpha\mu\nu\lambda}q^\nu Q^\lambda\right.\nonumber\\*
&& \mbox{} \left.
+ (\Delta^2 - q^2)\epsilon_{\alpha\beta\mu\nu} q^\nu
- 2i q_\alpha q_\beta (\tilde Q_\mu + \Delta \tilde \gamma_\mu)\gamma_5
\right]U^\beta\,,\nonumber\\
\bar U^{\prime\alpha}
\left\{R_\mu, P_\nu\right\}_{\alpha\beta} Q^\nu
U^\beta & = &
\bar U^{\prime\alpha}\left[
-\frac{i}{2}q^2 g_{\alpha\beta} \tilde Q_\mu \gamma_5
+ 2 q_\alpha \epsilon_{\beta\mu\nu\lambda} q^\nu Q^\lambda
+ 2 q_\beta \epsilon_{\alpha\mu\nu\lambda} q^\nu Q^\lambda\right.\nonumber\\
&&\mbox{} \left.
+ (\Delta^2 - q^2)\epsilon_{\alpha\beta\mu\nu} q^\nu
+ i(M^2 - q^2)
(q_\alpha \tilde g_{\beta\mu} - q_\beta \tilde g_{\alpha\mu})\gamma_5
\right]U^\beta\,.
\end{eqnarray}

These formulas allow us to determine by inspection the
relations between the set of form factors introduced in \Eq{ab}
and those introduced in \Eq{physicalparametrization}. Thus,
\begin{eqnarray}
\label{aAreln}
a_1 & = & -\frac{1}{2} M B_1 - \frac{7}{4} q^2 M D_1 \,,\nonumber \\ 
a^\prime_1 & = & -\frac{1}{2}i\Delta B_2 - \frac{1}{2} (q^2 - \Delta^2)M C_3 
- \frac{5}{4} iq^2\Delta E_1 - \frac{3}{8} iq^2 \Delta E_2 \,,\nonumber \\ 
a_2 & = & A_1 + \frac{1}{2} B_1 + \frac{7}{4} q^2 C_1
+ \frac{7}{4} q^2 D_1 \,,\nonumber \\
a^\prime_2 & = & iA_2 - \frac{1}{2}iB_2 - \frac{5}{4} i q^2 C_2
+ \frac{1}{2} M\Delta C_3 - \frac{i}{2} q^2 C_4
- \frac{i}{2} q^2 C_5 - \frac{5}{4} i q^2 E_1
- \frac{3}{8} iq^2 E_2 \,,\nonumber \\
a_3 & = & -4M D_1 \,,\nonumber \\
a^\prime_3 & = & -2i \Delta C_5 - 2i\Delta E_2 \,,\nonumber \\
a_4 & = & 4C_1  + 4 D_1 \,,\nonumber \\
a^\prime_4 & = & -2i C_5 - 2iE_2 \,,\nonumber \\
a_5 & = & 4iq^2 D_2 \,,\nonumber \\
a^\prime_5 & = & 0 \,,\nonumber \\
a_6 & = & -B_1 - \frac{15}{2} q^2 D_1  \,,\nonumber \\
a^\prime_6 & = & i (q^2 - M^2) C_4 + \frac{5}{2} i q^2 E_1 \,,\nonumber \\
b_1 & = & -B_2 - (q^2 - \Delta^2) (C_4 + C_5)
- \frac{3}{4} q^2 E_2\,,\nonumber \\
b_2 & = & -iC_3 + 2C_4 - C_5 \,,\nonumber \\
b_3 & = & iC_3 + 2C_4 - C_5 \,. 
\end{eqnarray}
Apart from the immediate result that $a^\prime_5 = 0$,
other less obvious relations that follow from the above formulas are
\begin{eqnarray}
\label{a34primerel}
a^\prime_3 & = & \Delta a^\prime_4 \,,\\
\label{a136rel}
\frac{1}{2}M a_6 & = & a_1 + \frac{1}{2}q^2 a_3 \,.
\end{eqnarray}
By considering \Eq{aAreln} in the diagonal case,
in which $\Delta = 0$, it is also straightforward to verify that
the conditions in \Eqs{qzeroconditions}{nonpollutionexamplerelations}
are indeed satisfied identically by these formulas. For example,
with $M = 2m$, \Eq{a136rel} precisely has the form stated in
\Eq{nonpollutionexamplerelations}.
In conclusion, the parametrization given in \Eq{physicalparametrization}
satisfies the conditions that the vertex function must satisfy, as
elaborated in Sections\ \ref{subsec:qzero} and \ref{subsec:pollution}.

\subsection{Multipole moment interactions}

Equation\ (\ref{physicalparametrization}) also has the virtue that it allows
us to identify the contributions from the various multipole moment terms
systematically. Since the correspondence between such terms and the
traditional definitions of the multipole moments using the static
and the nonrelativistic limit is not very useful
in the context of the applications that we envisage
(where neither limit may be applicable), we proceed in
a different direction. Making reference to the discussion in
Section\ \ref{subsec:nrlimit}, we write the amplitude $M$ in the presence
of an external potential in the form
\begin{equation}
M = -\bar U^{\prime\alpha} ({\cal M})_{\alpha\beta} U^\beta \,,
\end{equation}
where
\begin{equation}
({\cal M})_{\alpha\beta} = \Gamma_{\mu\alpha\beta} A^\mu(q) \,.
\end{equation}
Notice that, since we have constructed $\Gamma_{\mu\alpha\beta}$ such that
it is transverse, we can take $A^\mu(q)$ to be transverse since the
longitudinal part does not contribute. Introducing the field strength
\begin{equation}
F_{\mu\nu}(q) = i(q_\mu A_\nu(q) - q_\nu A_\mu(q))\,,
\end{equation}
we have the following relations
\begin{eqnarray}
R\cdot A(q) & = & \frac{i}{2}\Sigma_{\mu\nu} F^{\mu\nu} \nonumber\\
P\cdot A(q) & = & \frac{i}{2}\Sigma_{\mu\nu} \tilde F^{\mu\nu} = 
\frac{i}{2}\tilde\Sigma_{\mu\nu} F^{\mu\nu}\nonumber\\
q^2\tilde g_{\mu\nu} A^\nu(q) & = & iF_{\mu\nu}q ^\nu \,.
\end{eqnarray}
It is then evident that ${\cal M}$ has the following expansion in powers
of $q$,
\begin{eqnarray}
\label{multipoleexpansion}
{\cal M} = {\cal M}_0 +
{\cal M}_{\mu\nu} F^{\mu\nu}(q) +
i{\cal M}_{\lambda\mu\nu} q^\lambda F^{\mu\nu}(q) + 
{\cal M}_{\lambda\rho\mu\nu} q^\lambda q^\rho F^{\mu\nu}(q) +
i{\cal M}_{\lambda\sigma\tau\mu\nu} q^\lambda q^\sigma q^\tau F^{\mu\nu}(q)\,,
\end{eqnarray}
where, with the exception of ${\cal M}_0$, as noted below,
the various coefficients depend on $q$ only through the $q^2$
dependence of the form factors $A, B, C, D, E$. By a straightforward
calculation we find,
\begin{eqnarray}
\label{multipoleterms}
{\cal M}_0 & = & 
\left\{
\begin{array}{ll}
A_1 Q_\mu A^\mu(q) & \mbox{(charged diagonal case)}\\[12pt]
\frac{iA_1}{q^2} Q_\mu g_{\lambda\nu} q^\lambda F^{\mu\nu}(q)
& \mbox{(otherwise)}\\
\end{array}\right.\nonumber\\
{\cal M}_{\mu\nu} & = & -\frac{1}{2}(B_1\Sigma_{\mu\nu}
+ B_2\tilde \Sigma_{\mu\nu}) \nonumber\\
{\cal M}_{\lambda\mu\nu} & = & \frac{iA_2}{q^2} Q_\mu
g_{\lambda\nu} \gamma_5
- \frac{1}{2} \left(
iC_3 [\Sigma_{\mu\nu},\tilde\Sigma_{\lambda\rho}]
+ C_4 \{\Sigma_{\mu\nu},\tilde\Sigma_{\lambda\rho}\}
+ C_5 \{\tilde\Sigma_{\mu\nu},\Sigma_{\lambda\rho}\}\right) Q^\rho\,,
\nonumber\\
{\cal M}_{\lambda\rho\mu\nu} & = & -\frac{D_1}{4}
\left\{\Sigma_{\mu\nu},
\{\Sigma_{\sigma\lambda}, {\Sigma^\sigma}_\rho\}
\right\} +
\frac{iD_2}{4}\left[\Sigma_{\mu\nu},
\{\Sigma_{\sigma\lambda},{\Sigma^\sigma}_\rho\}\right]
\nonumber\\
&&\mbox{} - \frac{E_1}{4}\left\{\Sigma_{\mu\nu},\{\Sigma_{\sigma\lambda},
{\tilde\Sigma^\sigma}{}_\rho\}\right\} -
\frac{E_2}{4}\left(
\Sigma_{\sigma\lambda}\tilde\Sigma_{\mu\nu}{\Sigma^\sigma}_\rho +
(\lambda \leftrightarrow \rho)\right)\nonumber\\
{\cal M}_{\lambda\sigma\tau\mu\nu} & = & \frac{C_1}{2q^2}
\{\Sigma_{\rho\sigma}, {\Sigma^\rho}_{\tau}\}Q_\mu g_{\lambda\nu}
+ \frac{C_2}{2q^2}
\{\Sigma_{\rho\sigma}, {\tilde\Sigma^\rho}{}_\tau\} Q_\mu g_{\lambda\nu} \,,
\end{eqnarray}
where, in writing the $C_2$ and $E_1$ terms we have taken into
account \Eq{RPcommutator2}.

In the charged diagonal case, the term ${\cal M}_0$ is just the
electric charge interaction with the external potential. But
in the off-diagonal case, or in the diagonal case if the particle is
electrically neutral, so that $A_1(0) = 0$, the term in ${\cal M}_0$
actually represents an additional contribution to ${\cal M}_{\lambda\mu\nu}$
of the form
\begin{equation}
{\cal M}^\prime_{\lambda\mu\nu} = \frac{iA_1}{q^2} Q_\mu g_{\lambda\nu} \,.
\end{equation}
Equations\ (\ref{multipoleexpansion}) and (\ref{multipoleterms})
together represent a clear separation of the
various multipole moment interactions in terms of the form factors
defined in \Eq{physicalparametrization}.

\section{Implications of the discrete symmetries}
\label{s:discrete}

The form factors that appear in the vertex function may satisfy
additional requirements depending on whether the interaction Lagrangian
is invariant under the various discrete space-time symmetries,
and/or further conditions depending, for example, on whether
the particle is self-conjugate (Majorana condition) or not.
We consider the implications of all these conditions here.

Let us consider the $f(k) \rightarrow f^\prime(k^\prime) + \gamma$
transition amplitude where $f,f^\prime$ are RS fermions.
Using the definition of the vertex function in \Eq{jGammadef}
the amplitude is given by
\begin{equation}
\label{defM}
M(f(k) \rightarrow f^\prime(k^\prime) + \gamma) =
\bar U^{\prime\alpha}(k^\prime)\Gamma_{\alpha\beta\mu}(k,k^\prime)
U^\beta(k)\varepsilon^{\ast\mu} \,,
\end{equation}
where, in this section, we will consider the fermion momenta as the
independent variables in the vertex function.

First of all, independently of whether or not any of the
discrete transformations may be symmetries of the Lagrangian,
the hermiticity of the Lagrangian and
the substitution rule (crossing) yield the following two relations,
\begin{eqnarray}
\label{MH}
M(f^\prime(k^\prime) \rightarrow f(k) + \gamma) & = &
\bar U^{\beta}(k)
\bar\Gamma_{\alpha\beta\mu}(k,k^\prime)
U^{\prime\alpha}(k^\prime)\varepsilon^{\ast\mu} \,,\\
\label{Mcross}
M(\bar f^\prime(k^\prime) \rightarrow \bar f(k) + \gamma) & = &
\bar U^{\alpha}(k)
\Gamma^C_{\alpha\beta\mu}(-k,-k^\prime)
U^{\prime\beta}(k^\prime)\varepsilon^{\ast\mu} \,,
\end{eqnarray}
respectively, where
\begin{eqnarray}
\label{GammaBarandC}
\bar\Gamma_{\alpha\beta\mu}(k,k^\prime) & = &
\gamma^0 \Gamma^\dagger_{\alpha\beta\mu}(k,k^\prime)\gamma^0\,,\nonumber\\
\Gamma^C_{\alpha\beta\mu}(k,k^\prime) & = &
C^{-1} \Gamma^\top_{\beta\alpha\mu}(k,k^\prime)C\,,
\end{eqnarray}
with $C$ being the matrix with the property
$C^{-1}\gamma^\top_\mu C = -\gamma_\mu$ (e.g., $C = i\gamma_2\gamma_0$ in
the Dirac representation of the $\gamma$ matrices).
The vertex function $\Gamma^C_{\alpha\beta\mu}$ in \Eq{Mcross}
is obtained from $\Gamma_{\alpha\beta\mu}$,
by multiplying every quantity that
appears in $\Gamma_{\alpha\beta\mu}$ by its $C$ parity phase $\delta_C$ as
given in Tables\ \ref{tableCPT} and \ref{tableCPTprods}.
\begin{table}
\begin{displaymath}
\begin{array}{ccccccc}
\hline
& \delta_C & \delta_P & \delta_T & \delta_{CP} & \delta_{CPT} &
\delta_H \\ \hline
i & + & + & - & + & - & -\\
\gamma_5 & + & - & + & - & - & -\\
\gamma_\alpha & - & + & - & - & + & +\\
\gamma_\alpha\gamma_5 & + & - & - & - & + & +\\
\sigma_{\mu\nu} & - & + & - & - & + & +\\
\sigma_{\mu\nu}\gamma_5 & - & - & - & + & - & -\\
S_{\mu\nu} & - & + & - & - & + & +\\
\tilde S_{\mu\nu} & - & - & + & + & + & +\\
\Sigma_{\mu\nu} & - & + & - & - & + & +\\
\tilde\Sigma_{\mu\nu} & - & - & + & + & + & +
\end{array}
\end{displaymath}
\caption{Transformation rules under the discrete transformations
of the various quantities that appear in the vertex function.
\label{tableCPT}
}
\end{table}
\begin{table}
\begin{displaymath}
\begin{array}{ccccccc}
\hline
& \delta_C & \delta_P & \delta_T & \delta_{CP} & \delta_{CPT} &
\delta_H \\ \hline
\{\Sigma_{\mu\nu},\Sigma_{\lambda\rho}\}
        & + & + & + & + & + & +\\
\{\Sigma_{\mu\nu},\tilde\Sigma_{\lambda\rho}\}
        & + & - & - & - & + & +\\
{} [\Sigma_{\mu\nu},\tilde\Sigma_{\lambda\rho}]
        & - & - & - & + & - & -\\
\big\{\Sigma_{\mu\nu},\{\Sigma_{\lambda\rho},\Sigma_{\sigma\tau}\}\big\}
        & - & + & - & - & + & +\\
\big[\Sigma_{\mu\nu},\{\Sigma_{\lambda\rho},\Sigma_{\sigma\tau}\}\big]
        & + & + & - & + & - & -\\
\big\{
        \Sigma_{\mu\nu},\{\Sigma_{\lambda\rho},\tilde\Sigma_{\sigma\tau}\}
\big\}
        & - & - & + & + & + & +\\
\left(\Sigma_{\sigma\lambda}\tilde\Sigma_{\mu\nu}{\Sigma^\sigma}_\rho +
        (\lambda\leftrightarrow\rho)
\right)
         & - & - & + & + & + & +
\end{array}
\end{displaymath}
\caption{Transformation rules under the discrete transformations
of the various products of $\Sigma_{\mu\nu}$ and $\tilde\Sigma_{\mu\nu}$ 
that appear in the vertex function.
\label{tableCPTprods}}
\end{table}

We now envisage calculating the amplitude
using a transformed interaction Lagrangian that
is obtained from the original one by replacing each field by its
transformed counterpart. For example, for the parity transformation,
the photon and the RS fields would be replaced by
\begin{eqnarray}
A^\mu & \rightarrow & {\Lambda^\mu_P}_\nu A^\nu(\Lambda^{-1}_Px) \nonumber\\
\psi^\mu & \rightarrow & \eta_P {\Lambda^\mu_P}_\nu \gamma^0
\psi^\nu(\Lambda^{-1}_Px)\,,
\end{eqnarray}
where $\eta_P$ is a phase factor.
Other relevant fields are similarly replaced
by their corresponding parity-transformed counterpart.
Then, the one-photon transition amplitude obtained in this way,
which we denote by $M^P(f(k) \rightarrow f^\prime(k^\prime) + \gamma)$,
is given by
\begin{equation}
\label{defMP}
M^P(f(k) \rightarrow f^\prime(k^\prime) + \gamma) =
\eta_P \eta^{\prime\ast}_P
\bar U^{\prime\alpha}(k^\prime)\Gamma^P_{\alpha\beta\mu}(k,k^\prime)
U^\beta(k)\varepsilon^{\ast\mu} \,,
\end{equation}
where
\begin{equation}
\Gamma^P_{\alpha\beta\mu}(k,k^\prime) =
\gamma_0 \Gamma_{\alpha\beta\mu}(k,k^\prime)\gamma_0\,,
\end{equation}
and it is obtained from
$\Gamma_{\alpha\beta\mu}(k,k^\prime)$ by multiplying every quantity that
appears in $\Gamma_{\alpha\beta\mu}$ by its parity phase $\delta_P$ as
given in Tables\ \ref{tableCPT} and \ref{tableCPTprods}.

If the Lagrangian is invariant under the given transformation, then
the two amplitudes should be equal and therefore
\begin{equation}
\label{GammaPrel}
\Gamma^P_{\alpha\beta\mu}(k,k^\prime) =
\Gamma_{\alpha\beta\mu}(k,k^\prime) \,,
\end{equation}
which results in conditions on the form factors. Similarly,
for time-reversal and charge conjugation,
\begin{eqnarray}
\label{defMT}
M^T(f(k) \rightarrow f^\prime(k^\prime) + \gamma) & = &
-\eta_T \eta^{\prime\ast}_T\bar U^{\prime\alpha}(k^\prime)
\Gamma^T_{\alpha\beta\mu}(-k,-k^\prime)
U^\beta(k)\varepsilon^{\ast\mu} \,,\\
\label{defMC}
M^C(f^\prime(k^\prime) \rightarrow f(k) + \gamma) & = &
-\eta_C \eta^{\prime\ast}_C \bar U^{\alpha}(k)
\Gamma^C_{\alpha\beta\mu}(-k,-k^\prime)
U^{\prime\beta}(k^\prime)\varepsilon^{\ast\mu} \,,
\end{eqnarray}
respectively, where
\begin{equation}
\Gamma^T_{\alpha\beta\mu}(k,k^\prime) =
(\overline{C\gamma_5})\Gamma^\ast_{\alpha\beta\mu}(-k,-k^\prime)(C\gamma_5) \,,
\end{equation}
and $\Gamma^C_{\alpha\beta\mu}$ has been defined in \Eq{GammaBarandC}.
$\Gamma^T_{\alpha\beta\mu}(k,k^\prime)$ is obtained
from $\Gamma_{\alpha\beta\mu}(k,k^\prime)$
by multiplying every quantity that appears in
$\Gamma_{\alpha\beta\mu}(k,k^\prime)$ by its corresponding
$\delta_{T}$ phase.

We now consider separately the application of these relations to various
cases.

%
%
\begin{table}
\begin{displaymath}
\begin{array}{ccccccc}
\hline
& \delta_C & \delta_P & \delta_T & \delta_{CP} & \delta_{CPT} &
\delta_H \\ \hline
\tilde Q_\mu
        & - & + & - & - & + & + \\
i\tilde Q_\mu\gamma_5
        & - & - & + & + & + & + \\
iR_\mu
        & - & + & - & - & + & + \\
iP_\mu
        & - & - & + & + & + & + \\
i[R_\mu,P_\nu]Q^\nu
        & + & - & - & - & + & +\\
(R\cdot P)\tilde Q_\mu, \{R_\mu,P_\nu\}Q^\nu, \{P_\mu,R_\nu\}Q^\nu
        & - & - & + & + & + & +\\
i\{R_\mu,R\cdot R\}
        & - & + & - & - & + & +\\
{} [R_\mu, R\cdot R]
        & + & + & + & + & + & +\\
i\{R_\mu, R\cdot P\}, iR^\lambda P_\mu R_\lambda
        & - & - & + & + & + & +
\end{array}
\end{displaymath}
\caption{Transformation rules under the discrete transformations
of the various matrices that appear in the parametrization of
vertex function given in \Eq{physicalparametrization}. In addition, it must
be remembered that for $T$ and $H$ each form factor is replaced
by its complex conjugate.
\label{tableCPTRPQ}
}
\end{table}

\subsection{Diagonal case $f = f^\prime$}
\label{sec:diracdiagcase}

In this case, we can summarize the effect of performing the discrete
transformations $C$, $P$ and $T$ by saying that
the amplitude for the process
$f(k) \rightarrow f(k^\prime) + \gamma$,
calculated with the transformed Lagrangian, is obtained by
making the substitutions
\begin{eqnarray}
\Gamma_{\alpha\beta\mu}(k,k^\prime) & \stackrel{C}{\rightarrow} &
-\Gamma^C_{\alpha\beta\mu}(-k^\prime,-k) \,,\nonumber\\
\Gamma_{\alpha\beta\mu}(k,k^\prime) & \stackrel{P}{\rightarrow} &
\Gamma^P_{\alpha\beta\mu}(k,k^\prime) \,,\nonumber\\
\Gamma_{\alpha\beta\mu}(k,k^\prime) & \stackrel{T}{\rightarrow} &
-\Gamma^T_{\alpha\beta\mu}(-k,-k^\prime) \,.
\end{eqnarray}
From these, the transformation rules for the combined transformations
follow, for example,
\begin{equation}
\Gamma_{\alpha\beta\mu}(k,k^\prime) \stackrel{CP}{\rightarrow} 
-\Gamma^{CP}_{\alpha\beta\mu}(-k^\prime,-k) \,.
\end{equation}
If the Lagrangian is invariant under a given transformation,
then the arrow symbol in the corresponding relation is replaced
by the equals sign, which results in conditions on the form factors.

In order to ease the application of such conditions to the
parametrization given in \Eq{physicalparametrization},
the transformation rules of the various terms that appear in that equation
are displayed in Table\ \ref{tableCPTRPQ}.
For example, starting from \Eq{physicalparametrization}, the expression
for $\Gamma^{CP}_{\alpha\beta\mu}(-k^\prime,-k)$ is obtained by
multiplying every term in that equation by its corresponding phase
$\delta_{CP}$ as given in Table\ \ref{tableCPTRPQ}.

As an example, let us consider $CP$. If the relevant interaction
Lagrangian is $CP$ symmetric, then only the terms with $\delta_{CP}$ odd
in Table\ \ref{tableCPTRPQ} can appear in the vertex function.
Therefore, only $A_1$, $B_1$, $C_3$ and $D_1$ may be nonzero in that case.

Independently of any such conditions, \Eq{MH} implies that the
vertex function in this case satisfies
\begin{equation}
\label{diaghermiticity}
\Gamma^H_{\alpha\beta\mu}(k,k^\prime) = \Gamma_{\alpha\beta\mu}(k,k^\prime)\,,
\end{equation}
where $\Gamma^H_{\alpha\beta\mu}$ is defined by
\begin{equation}
\Gamma^H_{\alpha\beta\mu}(k,k^\prime) \equiv
\bar\Gamma_{\beta\alpha\mu}(k^\prime,k) \,,
\end{equation}
with $\bar\Gamma_{\alpha\beta\mu}(k,k^\prime)$ defined
in \Eq{GammaBarandC}. $\Gamma^H_{\alpha\beta\mu}(k,k^\prime)$
is obtained from $\Gamma_{\alpha\beta\mu}(k,k^\prime)$
in \Eq{physicalparametrization} by multiplying every quantity that
appears there by the phase $\delta_H$
given in Table\ \ref{tableCPTRPQ} (and replacing each form factor
by its complex conjugate). Thus, \Eq{diaghermiticity} implies that
all the form factors $A, B, C, D , E$ in \Eq{physicalparametrization} are real. 

\subsubsection{Self-conjugate (Majorana) particles}

In this case, \Eq{Mcross} implies the additional relation
\begin{equation}
\label{MajoranaGammaRel}
\Gamma_{\alpha\beta\mu}(k,k^\prime) = 
\Gamma^C_{\alpha\beta\mu}(-k^\prime,-k) \,.
\end{equation}
irrespectively of the discrete symmetries that may exist. As can be seen
from Table\ \ref{tableCPTRPQ} the vertex function consists only
of the $C_3$ and $D_2$ terms. In terms of the parametrization defined
in \Eq{ab}, \Eq{aAreln} then implies that the vertex function in this
case is of the form
\begin{eqnarray}
\label{majdiagvertex}
\Gamma_{\alpha\beta\mu} & = &
- q^2 m C_3 g_{\alpha\beta} \tilde \gamma_\mu \gamma_5 +
i C_3 \left(
q_\alpha\epsilon_{\beta\mu\nu\lambda} q^\nu Q^\lambda -
q_\beta\epsilon_{\alpha\mu\nu\lambda} q^\nu Q^\lambda\right) \nonumber\\
&&\mbox{} + 4i q^2 D_2
(\tilde g_{\mu\alpha} q_{\beta} + \tilde g_{\mu\beta} q_{\alpha}) \,,
\end{eqnarray}
where we have used the fact that $m^\prime = m$ in the present case.

The term proportional to $\gamma_\mu\gamma_5$, which is related to the axial
charge radius, is reminiscent of the analogous result
that holds for Majorana neutrinos\cite{n:nuem}.
The other two terms in \Eq{majdiagvertex}
resemble the result that was obtained in Ref.\ \cite{np:spin1}
for the electromagnetic vertex function of self-conjugate spin-1
particles [Eq. (4.14) in that reference]. The noteworthy feature
of the result obtained here for the spin 3/2 case is the fact
that the coefficients of the first two terms in \Eq{majdiagvertex} above
are given in terms of the same form factor.
Ultimately this is related to the requirement that the vertex function
acting on the initial spinor, or on the final spinor,
does not yield a spinor with a spin 1/2 component as we have discussed
in the previous section, which the parametrization in terms of $C_3$ and $D_2$
automatically satisfies.

While the above result is general, further conditions exist if
some discrete symmetries hold. For example, if $CP$ holds,
then $D_2 = 0$ as already stated in Section\ \ref{sec:diracdiagcase},
so that only the $C_3$ terms in \Eq{majdiagvertex} can be present.

\subsection{Off-diagonal case $f \not= f^\prime$}

In the off-diagonal case, both the hermiticity condition and crossing
only relate the amplitude of one process to the amplitude
of a different process and therefore they do not lead by themselves
to any restriction on the form factors. The restrictions arise only
if some discrete symmetries hold. For example, if $CP$ is valid,
then we obtain the relation
\begin{equation}
\label{offdiagCP}
-\eta_{CP}\eta^{\prime\ast}_{CP} \Gamma^{CP}_{\alpha\beta\mu}(-k,-k^\prime) =
\Gamma^H_{\alpha\beta\mu}(k,k^\prime) \,,
\end{equation}
where $\eta_{CP}$ and $\eta^\prime_{CP}$ are the phases in the
$CP$ transformation rule of the fermion fields.
This implies that the group of form factors that appear in
\Eq{physicalparametrization} are divided in two groups. The first
group consists of
\begin{equation}
A_1, B_1, C_3, D_1 \,,
\end{equation}
which are the coefficients of the terms with $\delta_{CP}$ odd according
to Table\ \ref{tableCPTRPQ}, and therefore must be relatively real.
The second group consists of the remaining coefficients, which
are associated with the terms that have $\delta_{CP}$ even
in Table\ \ref{tableCPTRPQ}, and therefore must be relatively imaginary
with respect to the coefficients included in the first group.

\subsubsection{Self-conjugate (Majorana) particles}

In this case, \Eqs{MH}{Mcross} together imply that
\begin{equation}
\Gamma^{C}_{\alpha\beta\mu}(-k,-k^\prime) =
\Gamma^H_{\alpha\beta\mu}(k,k^\prime) \,,
\end{equation}
independently of the discrete symmetries. Thus, the form factors are again
divided in two groups,
\begin{equation}
\label{offdiagmajorana}
C_3, D_2 = \mbox{real}
\end{equation}
while all the others are imaginary.

The discrete symmetries would yield additional conditions.
Considering $CP$ as an example once more, if it holds, then
the same argument that lead to \Eq{offdiagCP}, in this case leads to
\begin{equation}
-\eta\Gamma^{CP}_{\alpha\beta\mu}(-k,-k^\prime) =
\Gamma^H_{\alpha\beta\mu}(k,k^\prime) \,.
\end{equation}
where $\eta$ is $+1$ or $-1$ according to whether the two Majorana fermions
have the same or opposite $CP$ parity, respectively, Thus, if
the two fermions have the same $CP$ parity ($\eta = +1$), then
\begin{equation}
\label{offdiagmajoranaCP}
A_1, B_1, C_3, D_1 = \mbox{real}\,,
\end{equation}
while the other coefficients are imaginary. Thus,
\Eqs{offdiagmajorana}{offdiagmajoranaCP} together imply that in
this case only the $C_3$ is nonzero. Conversely, if the fermions
have the opposite $CP$ parity, then only $D_2$ is nonzero.
Once more, this is reminiscent of the analogous result for Majorana
neutrinos\cite{n:nuem} in which the transition electromagnetic moment
is of the form $\sigma_{\mu\nu}\gamma_5$ of $\sigma_{\mu\nu}$ when
the neutrinos have the same or opposite $CP$ phase, respectively.

\section{Discussion}
\label{s:discussion}

Although we have focused our attention on the Majorana case,
our results have application to the charged cases as well.
Let us consider the diagonal vertex function, assuming exact
$P$ and $CP$ symmetry as would be applicable to a baryon for example.
Since, as we have already seen, $CP$ in the diagonal case implies
that only $A_1$, $B_1$, $D_1$ and $C_3$ are non-zero, the additional
$P$ symmetry implies that only the first of those three are non-zero.
The vertex function then reduces to the form
\begin{equation}
\Gamma_{\alpha\beta\mu} =
g_{\alpha\beta} \gamma_\mu a_1 
+ g_{\alpha\beta} Q_\mu a_2
+ q_\alpha q_{\beta} \gamma_\mu a_3
+ q_\alpha q_{\beta} Q_\mu a_4
+ (g_{\mu\alpha} q_{\beta} - g_{\mu\beta} q_{\alpha}) a_6\,,
\end{equation}
where, remembering that $M = 2m$ in this case,
\begin{eqnarray}
a_1 & = & -m B_1 - \frac{7}{2} q^2 m D_1 \,,\nonumber\\
a_2 & = & A_1 + \frac{1}{2} B_1 + \frac{7}{4} q^2 D_1 \,,\nonumber \\
a_3 & = & -8 m D_1 \,,\nonumber \\
a_4 & = & 4 D_1 \,,\nonumber \\
a_6 & = & -B_1 - \frac{15}{2} q^2 D_1 \,.
\end{eqnarray}
Of course we can choose to write the vertex
function in terms of any three independent parameters. Choosing
$a_{1,2,3}$, so that
\begin{eqnarray}
a_4 & = & -\frac{1}{2m} a_3 \,,\nonumber\\
a_6 & = & \frac{1}{m}\left(a_1 + \frac{1}{2} q^2 a_3\right) \,,
\end{eqnarray}
then yields
\begin{equation}
\label{minimalcoupling}
\Gamma_{\alpha\beta\mu} =
\left[g_{\alpha\beta} \gamma_\mu -
\frac{i}{m} (S_{\mu\nu}q^\nu)_{\alpha\beta}\right] a_1
+ g_{\alpha\beta} Q_\mu a_2
+ q_\alpha q_{\beta}\left[\gamma_\mu - \frac{1}{2m}Q_\mu\right] a_3\,,
\end{equation}
where the $S_{\mu\nu}$ are the spin-1 generators of the Lorentz group
defined in \Eq{spin1generators}.

The function with the lowest powers of $q$, such as what might arise from a
tree-level vertex, corresponds to setting $a_3 = 0$. In any case,
the terms with $Q_\mu$ could be rewritten using the Gordon identities
given in \Eq{Gordon}, if desired. But regardless of that, this equation
reveals that the electromagnetic vertex in the RS representation
cannot consist solely of the $g_{\alpha\beta}\gamma_\mu $ term.
The underlying reason is that while each of the couplings
$g_{\alpha\beta}Q_\mu$ and $(\Sigma_{\mu\nu} q^\nu)_{\alpha\beta}$
independently satisfy a relation similar to \Eq{nsigSigrelationCov}),
and therefore the vertex can contain one or the other or both, the couplings
$\gamma_\mu g_{\alpha\beta}$ and $\sigma_{\mu\nu} q^\nu g_{\alpha\beta}$
do not satisfy any such relations. Thus in contrast to the spin-1/2 case,
in the spin-3/2 case the $\gamma_\mu$ coupling term must be
accompanied by the appropriate spin coupling term, as shown in
\Eq{minimalcoupling}, exposing the fact that such is the combination
that is actually hidden inside the $Q_\mu g_{\alpha\beta}$ and/or
$\Sigma_{\mu\nu} q^\nu$ couplings. This indicates that
the minimal substitution prescription in the RS representation,
which leads to a pure $\gamma_\mu$ coupling at the tree-level,
can lead to inconsistencies
(e.g., nonunitarity problems\cite{delgado1,delgado2})
related to the nondecoupling of the spurious spin-1/2 components
of the RS representation.

\section{Conclusions}

We have studied the structure of the electromagnetic vertex function of
spin-3/2 particles using the Rarita-Schwinger description of such particles,
including both the diagonal and off-diagonal (transition) cases.
We considered the cases in which the particles
are charged or electrically neutral as well as the particular case
that they are self-conjugate (Majorana type).

In Section\ \ref{s:generalform} we considered the general form
of the on-shell vertex function that is consistent with electromagnetic
gauge invariance in terms of simple combinations of the $\gamma$ matrices,
the momentum vectors, the metric and the Levi-Civita tensors.
While that representation is simple and useful for practical calculations
of transition rates, it is not practical for taking into account
the physical requirement that the vertex function does not mix
the genuine spin-3/2 degrees of freedom with the spurious spin-1/2
components of the RS representation.
Thus, in Section\ \ref{s:physicalparametrization}
we considered another expression for the on-shell vertex function,
in terms of the matrices $R_\mu, P_\mu$, and their products.
We showed that, without having to impose further conditions, this
construction of the vertex function satisfies the physical requirements
related to gauge invariance and the use of the RS spinors, avoiding
the spin-1/2 - spin-3/2 mixing problem to which we have alluded above.
The formulas that give the relations between the form factors of the two
representations were given explicitly there as well. Finally in
Section\ \ref{s:discrete} we studied the implications due to the discrete
transformations, such as the $C, P, T$ transformations and their products, and
the conditions implied by the hermiticity of the interaction Lagrangian
and crossing symmetry, for both the diagonal and off-diagonal cases.
We considered in some detail the diagonal Majorana case,
and showed that the vertex function can contain
a term of the form $\gamma_\mu\gamma_5$,
which resembles the axial charge radius term for Majorana
neutrinos\cite{n:nuem}, plus another one that resembles the vertex
function for self-conjugate spin-1 particles\cite{np:spin1},
but with the particularity that the two terms appear
with a specific relative coefficient and not independently.
In essence this result is due to the requirement that
the vertex function does not mix the genuine spin-3/2 degrees of freedom
with the spurious spin-1/2 components of the RS representation,
which the expression for the vertex function in terms of the matrices
$R_\mu, P_\mu$ automatically satisfies.
The analogous results for the off-diagonal vertex, as well as the
charged particles cases, were discussed as well.

The analysis that we have presented can be a useful
in several ways.  On one hand, it can serve as a guide to parametrize
the electromagnetic couplings of spin-3/2 particles
in a way that is general, model-independent and consistent.
On the other hand, the results of our analysis can be used to test deviations
from fundamental physical principles, such as gauge invariance and
crossing symmetry, in the context of the processes described by the
electromagnetic couplings of such particles, in particular Majorana ones.
In addition it can serve as a guide to study the
spin-3/2 to spin-1/2 transition vertex, for example the gravitino-neutrino
radiative transition which can be of interest in cosmological contexts.
\appendix
\section{Reduction formulas for products of $R$ and $P$} 

%
%
%
\Eqs{RR}{RP} can in turn be used to find the expressions for the
terms in \Eq{physicalparametrization} that involve three
factors of $R$, $P$ and $Q$,
\begin{eqnarray}
\label{cubicterms}
%
%
(R_\mu R\cdot R)_{\alpha\beta} & = &
\frac{7}{4} q^2 (R_\mu)_{\alpha\beta} +
2 q_\alpha q_\beta\sigma_{\mu\nu} q^\nu +
4i\left(q^2 g_{\mu\alpha} - q_\mu q_\alpha\right)q_\beta
+ (R_\mu L_1)_{\alpha\beta}\,,\nonumber\\ 
%
%
(R\cdot R R_\mu)_{\alpha\beta} & = &
\frac{7}{4} q^2 (R_\mu)_{\alpha\beta} +
2 q_\alpha q_\beta\sigma_{\mu\nu} q^\nu -
4i q_\alpha \left(q^2 g_{\mu\beta} - q_\mu q_\beta \right)
+ (L_1 R_\mu)_{\alpha\beta}\,,\nonumber\\ 
%
%
(R_\mu R\cdot P)_{\alpha\beta} & = &
-\frac{5}{4}i q^2 (R_\mu)_{\alpha\beta}\gamma_5 + (R_\mu L_2)_{\alpha\beta}
\,,\nonumber\\
%
%
(R\cdot P R_\mu)_{\alpha\beta} & = &
-\frac{5}{4}i q^2 (R_\mu)_{\alpha\beta}\gamma_5 + (L_2 R_\mu)_{\alpha\beta}
\,,\nonumber\\
%
%
(R^\lambda P_\mu R_\lambda)_{\alpha\beta} & = &
\frac{3}{4} q^2 (P_\mu)_{\alpha\beta}
- 2i q_\alpha q_\beta \sigma_{\mu\nu} \gamma_5 q^\nu
+ (P_\mu L_1)_{\alpha\beta}
\end{eqnarray}
The terms that multiply $C_{3,4,5,6}$ in
\Eq{physicalparametrization} can be expressed explicitly in similar
fashion,
\begin{eqnarray}
\label{RP:}
\left[R_\mu, P_\nu\right]_{\alpha\beta} & = &
q_\beta\epsilon_{\mu\nu\alpha\lambda} q^\lambda
- q_\alpha\epsilon_{\mu\nu\beta\lambda} q^\lambda
- \frac{1}{2}\Big(q^2 {\tilde g_{\mu}}^\lambda\sigma_{\lambda\nu}
- q_\nu\sigma_{\mu\lambda}q ^\lambda\Big) g_{\alpha\beta}\gamma_5
\,,\nonumber \\
\left\{R_\mu, P_\nu\right\}_{\alpha\beta} & = &
-\frac{i}{2}q^2\tilde g_{\mu\nu} g_{\alpha\beta}\gamma_5
+ i\epsilon_{\nu\alpha\beta\rho}q ^\rho\sigma_{\mu\lambda} q^\lambda
\,,\nonumber\\
&& \mbox{} + (q_\beta g_{\mu\alpha} - q_\alpha g_{\mu\beta})
\sigma_{\nu\lambda}q ^\lambda\gamma_5
- q_\beta\epsilon_{\mu\nu\alpha\lambda} q^\lambda
- q_\alpha\epsilon_{\mu\nu\beta\lambda} q^\lambda \,,
\end{eqnarray}
where use has been made of the commutator of the $\sigma_{\mu\nu}$
as well as their anticommutator
\begin{eqnarray}
\{\sigma_{\mu\nu},\sigma_{\lambda\rho}\} = 2
(g_{\mu\lambda} g_{\nu\rho} - g_{\mu\rho} g_{\nu\lambda})
+ 2i\epsilon_{\mu\nu\lambda\rho} \gamma_5 \,,
\end{eqnarray}
which in particular yields
\begin{equation}
\{\sigma_{\mu\nu}q^\nu ,\sigma_{\lambda\rho}q^\rho\} =
2q^2 \tilde g_{\mu\lambda} \,.
\end{equation}

In obtaining some of these formulas, we have found useful the
following expression for $P_\mu$,
\begin{equation}
\label{equivalentPform}
(P_\mu)_{\alpha\beta} = i\left(\epsilon_{\mu\nu\alpha\beta}
- \frac12 g_{\alpha\beta}\sigma_{\mu\nu}\gamma_5\right) q^\nu \,.
\end{equation}
For example, this easily gives the relation
\begin{equation}
q^\alpha (P_\mu)_{\alpha\beta} = - \frac12 i\sigma_{\mu\nu}
\gamma_5 q^\nu q_\beta \,.
\end{equation}
The latter formula, together with the relation
\begin{equation}
\lslash{q}\sigma_{\mu\nu}\gamma_5 = \sigma_{\mu\nu}\gamma_5\lslash{q}
\end{equation}
and
\begin{equation}
\gamma^\alpha (P_\mu)_{\alpha\beta} =
- \frac12 i \sigma_{\mu\nu} \gamma_5 q^\nu \gamma_\beta \,,
\end{equation}
which follows from \Eq{nsigSigrelationCov}, allows us
to confirm explicitly that
\begin{equation}
(L_1 P_\mu)_{\alpha\beta} = (P_\mu L_1)_{\alpha\beta} \,,
\end{equation}
which in turn implies that
$(R\cdot R P_\mu)_{\alpha\beta} = (P_\mu R\cdot R)_{\alpha\beta}$
as stated in \Eq{R2Prelations}.

In obtaining the last formula quoted in \Eq{RPcommanticommreduced}
we have used the Gordon identity to write
\begin{eqnarray*}
\bar U^{\prime\alpha}
i\epsilon_{\nu\alpha\beta\rho} q^\rho \sigma_{\mu\lambda} q^\lambda Q^\nu
U^\beta & = & \bar U^{\prime\alpha}\left[
- \epsilon_{\alpha\beta\lambda\rho} q^\lambda Q^\rho (Q_\mu - M\gamma_\mu)
\right]U^\beta
\end{eqnarray*}
which, using the identity in \Eq{5eidentity} and proceeding as indicated
there, can be reduced further to the form
\begin{eqnarray*}
\bar U^{\prime\alpha}
i\epsilon_{\nu\alpha\beta\rho} q^\rho \sigma_{\mu\lambda} q^\lambda Q^\nu
U^\beta & = &
\bar U^{\prime\alpha}\left[
q_\alpha \epsilon_{\beta\mu\nu\lambda} q^\nu Q^\lambda
+ q_\beta \epsilon_{\alpha\mu\nu\lambda} q^\nu Q^\lambda
+ (\Delta^2 - q^2)\epsilon_{\alpha\beta\mu\nu} q^\nu
\right]U^\beta\,.
\end{eqnarray*}

\end{document}